\documentclass[12pt, draftclsnofoot, onecolumn, journal]{IEEEtran}
\usepackage{amsmath,amssymb,pst-all,pseudocode}
\usepackage{epsfig,subfigure,color,graphicx}
\usepackage{pstricks}
\usepackage{pstricks-add}

\newtheorem{lemma}{Lemma}
\newtheorem{theorem}{Theorem}
\newtheorem{corollary}{Corollary}

\newtheorem{example}{Example}
\newtheorem{proposition}{Proposition}
\newcommand{\chsfk}[2]{\genfrac{[}{]}{0pt}{0}{#1}{#2}}

\newcommand{\rank}{\text{rank}}
\def\blfootnote{\xdef\@thefnmark{}\@footnotetext}

\begin{document}


\title{Secure Network Coding\\[-3mm] for  Wiretap Networks of Type II}
%\title{Secure Network Coding in the Presence of a Wiretapper}
%\title{Secure Coding for Wiretap Networks}
%alex - changed title


\author{Salim~ El~Rouayheb,
        Emina~Soljanin,
        Alex~Sprintson
%\author{
%\authorblockN{Salim Y. El Rouayheb}
%\authorblockA{ECE Department \\
%Texas A\&M University \\
%College Station, TX 77843 \\
%salim@ece.tamu.edu} \and
%\authorblockN{Emina Soljanin}
%\authorblockA{Math. Sciences Center \\
%Bell Labs, Alcatel-Lucent \\
%Murray Hill, NJ 07974 \\
%emina@alcatel-lucent.com}
%\and
%\authorblockN{Alex Sprintson}
%\authorblockA{ECE Department \\
%Texas A\&M University \\
%College Station, TX 77843 \\
%spalex@tamu.edu}
\thanks{S. El~Rouayheb and A. Sprintson are with the Department of Electrical and Computer Engineering, Texas A\&M University, College Station, TX, 77845 USA emails:~\{rouayheb@tamu.edu, spalex@ece.tamu.edu\}.}%
\thanks{E. Soljanin is with the Mathematics of Networking and Communication Deptartment, Enabling Computing Technologies, Bell Laboratories,  Alcatel-Lucent, Murray Hill, NJ 07974, email: emina@alcatel-lucent.com.}
\thanks{A preliminary version of this paper appeared in the Proceedings of the IEEE International Symposium on Information Theory (ISIT), Nice, France, 2007 \cite{RS07}.}
%Some of the results in this paper were presented  at the IEEE International %Symposium on Information Theory (ISIT), Nice, France, 2007 \cite{RS07}.}
}
\maketitle

\begin{abstract}
We consider the problem of securing a multicast network against a
wiretapper that can intercept the packets on a limited number
of arbitrary network edges of its choice. We assume that the network employs the network coding technique to simultaneously deliver the packets available at the source to all the receivers.
 We show that this problem can be looked at as a
network generalization of the  wiretap channel of type~II introduced in a seminal paper by Ozarow and Wyner.
In particular, we show that the transmitted information can be secured by using
the Ozarow-Wyner approach of coset coding at the source on top of the existing
network code. This way, we quickly and transparently recover some of the
results available in the literature on secure network coding for wiretap networks. Moreover,
we  derive new bounds on the required alphabet size that are independent of the network size  and  devise an algorithm for the construction of  secure network codes.
We also look at the dual problem and analyze
 the amount of information that  can be gained by the wiretapper as a function of the number of wiretapped edges.
\end{abstract}

\section{Introduction}
Consider a communication network represented as a directed graph
$G=(V,E)$ with unit capacity edges and an information source $S$
that multicasts information to $t$ receivers $R_1,\dots,R_t$ located at distinct nodes.  Assume that
the minimum size of a cut that separates the source and each receiver node is $n$. It is known that a multicast rate of $n$ is
achievable by using a linear network coding scheme \cite{Ahl, Li}.
In this paper, we focus on secure multicast connections in the presence of a wiretapper that can
access data on a limited number of  edges of its choice. Our primary goal is to design a network coding scheme that delivers data at maximum rate to all the destinations and does not reveal any information about the transmitted message to the wiretapper.

The problem of making a linear network code information-theoretically secure in the presence of a wiretaper that can
look at a bounded number, say $\mu$, of network edges was first
studied by Cai and Yeung in \cite{yeung02secure}. They considered
directed graphs and constructed codes over an
alphabet with at least $\binom{|E|}{\mu}$ elements which can support
a secure multicast rate of up to $n - \mu$.  In \cite{YC08}, they proved that these codes use the minimum amount of randomness required to achieve the security constraint.  However, the algorithm due to \cite{yeung02secure} has high computational complexity and requires a very large field size (exponential in the number of wiretapped edges).
%, and  later proved in \cite{YC08} their optimality in the sense of using the minimum amount of  randomness needed to achieve the security constraint.
%The high required field size and the code design complexity
%are the main drawbacks of this pioneering work.
Feldman \emph{et al.}
derived trade-offs between security, code alphabet size, and
multicast rate of secure linear network coding schemes in
\cite{feldman04csnc}, by using ideas from secret sharing and
abstracting the network topology. Another approach was taken by Jain in
\cite{jain04} who obtained security by merely exploiting the
topology of the underlying network. Weakly secure network  codes
%coding
that insure that no meaningful  information   is
revealed to the adversary were studied by Bhattad and Narayanan in
\cite{bhattad05secure}.
% and practical schemes are missing in this
%case as well.

A related line of work considers a more powerful \emph{Byzantine} adversary that can also modify the packets on the   edges it controls.
Such an adversary can be potentially more harmful in networks that employ the network coding technique because a modification in one packet can propagate throughout the network and affect other packets as well.
Secure network coding in the presence of a Byzantine adversary has been studied by Ho \emph{et al.} in \cite{ho04byzantine} and Jaggi \emph{et al.} in \cite{JLHE05, jaggi07infocom, JLKHKMM08}. In \cite{jaggi07infocom, JLKHKMM08}, the authors devise distributed polynomial-time algorithms that are rate-optimal and achieve information theoretical security  against  several
%various
scenarios of  adversarial attacks.

The problem of error correction in networks was also studied by Cai and Yeung in \cite{CY1,CY2} where they  generalized classical error-correction coding techniques to network settings. A different model  for error correction was introduced by Koetter and Kschischang in \cite{KK08} where communication is established by transmitting subspaces instead of vectors through the network.
 The use of rank-metric codes for error control under this model was investigated in \cite{SKK08}.
The common approach in these works is to encode packets at the source, prior to sending them over the network,  using an error correcting code so that the packets carry not only data but also some redundant information
derived from the data
which will help  to reduce the probability of incorrect decoding.

We also consider the coding at the source technique to be a natural approach for addressing the
information-theoretic security of wiretap networks. In a network
where the min-cut value between the source and each receiver node is
$n$ and an adversary can access up to $\mu$ edges of his choice, we
introduce  a coding at source scheme that ensures information-theoretic security based on the Ozarow-Wyner wiretap channel of type~II,
introduced in  \cite{Ozarow&Wyner:84} and \cite{OzarowWy85}, where
the source transmits $n$ symbols to the receiver and an adversary
can access any $\mu$ of those symbols.

Ozarow and Wyner showed that the maximum number of symbols (say $k$)
that the source can communicate to the receiver securely in the
information-theoretic sense is equal to $n-\mu$. They also showed
how to encode the $k$ source symbols into the $n$ channel symbols
for secure transmission. Clearly, if the $n$ channel symbols are
multicast over a network using a routing scheme, the $k$ source symbols remain secure in the
presence of an adversary with access to any $\mu$ edges. We will
illustrate later that this is not necessarily the case when
network coding is  used.
%performed.
 However, we will show that a network code based on the Ozarow-Wyner scheme
that preserves security of the $k$ source symbols, which are coded into the $n$ multicast symbols,
can be designed over a sufficiently large field.

Using the observations made by Feldman \emph{et al.}\ in
\cite{feldman04csnc}, we show that our scheme is
equivalent to the one proposed in the pioneering work of Cai and
Yeung in \cite{yeung02secure}. However, with our approach, we can
quickly and transparently recover some of the results available in
the literature on secure network coding for wiretapped networks.
The algorithm due to \cite{yeung02secure} is based on the code construction proposed by Li \emph{et al.}\ in \cite{Li}, however more efficient network coding algorithms have been proposed recently (see, e.g., \cite{jaggi03polynomial} and \cite{ceitnc}).
%
%Since the publication of \cite{yeung02secure} in which the network code construction is based on the work of Li \emph{et al.}\ in \cite{Li}, a number of simpler network code construction algorithms have been proposed (see for example \cite{jaggi03polynomial} and \cite{ceitnc}).
 We  use the results on the encoding complexity of the network coding  presented in
\cite{ceitnc}, \cite{lang06com}, \cite{lang06} to derive new bounds on the required field size of a secure network code that are independent of the number of edges in the network and that depend only on the number $k$ of source symbols and the number $t$ of destinations. We also propose an algorithm for construction of a secure network code that achieves these bounds. Furthermore, we look at the dual problem and analyze the
security of a given Ozarow-Wyner code
%performance of a given Ozarow-Wyner code in achieving security for  networks
by studying the amount of information that can be gained by the wiretapper as a function of the number of wiretapped edges.

 Parts of the results presented in this paper were published in \cite{RS07} and were later extended in \cite{SK08, SK08j} by Silva and Kschischang to construct universal secure network codes based on maximum rank-distance (MRD) codes, and by Mills \emph{et al.}  in \cite{MSCSV08} to achieve secrecy for wireless erasure networks.

This paper is organized as follows: In  Section~\ref{sec:wtc}, we briefly review the Ozarow-Wyner wiretap channel  of type II problem.
In Section~\ref{sec:wtn}, we introduce the network generalization of this problem.
In Section~\ref{sec:cd}, we present an algorithm for secure network code design and  establish new bounds on the required code alphabet size. In Section~\ref{sec:WiretapperEquivocation}, we study the   security  of Ozarow-Wyner codes. In Section~\ref{sec:con}, we highlight some connections of this work with other works on secure network coding and  network error correction. Finally, we conclude in Section~\ref{sec:conc} with a summary of our results  and open problems.

\section{Wiretap Channel II\label{sec:wtc}}
We first consider a point-to-point scenario in which the source can
transmit $n$ symbols to the receiver and an adversary can access any
$\mu$ of those symbols \cite{Ozarow&Wyner:84,OzarowWy85}. For this
case, we know that the maximum number of symbols that the source can
communicate to the receiver securely in the information-theoretic
sense is equal to $n-\mu$.

The problem is mathematically formulated as follows. Let
$S=(s_1,s_2,\dots,s_k)^T$ be the random variable  associated with the
%alex Salim - you need to be consistent - use variable or vector everywher
$k$ information symbols that the source wishes to send securely,
$Y=(y_1,y_2,\dots,y_n)^T$ the random variable associated with the symbols
that are transmitted through the noiseless channel between the
source and the receiver, and $Z=(z_1,z_2,\dots,z_{\mu})^T$ the random
variable associated with the wiretapped symbolsof $Y$. When $k\le
n-\mu$, there exists an encoding scheme that maps $S$ into $Y$ such that:
\begin{enumerate}
  \item The uncertainty about $S$ is not reduced by the knowledge of
$Z$ (perfect secrecy condition), \emph{i.e.}, \begin{equation}
H(S|Z)=H(S),  \label{eq:seccon}
\end{equation} and,
  \item  The information $S$ is completely determined (decodable) by the complete knowledge of $Y$, that is,
\end{enumerate}
\begin{equation}
H(S|Y)=0. \label{eq:deccon}
\end{equation}

For $n=2$, $k=1$, $\mu=1$, such a coding scheme can be constructed as
follows. If the source bit equals $0$, then either $00$ or $11$ is
transmitted through the channel with equal probability. Similarly,
if the source bit equals $1$, then either $01$ or $10$ is
transmitted through the channel with equal probability:
\[
 \begin{array}{|l| c|c| }
	\hline
 \text{source bit $s_1$} &  0 & 1\\
\hline
\text{codeword $y_1y_2$ chosen} & &\\ \text{at random from} & \{00, 11\} & \{01,
10\}\\
\hline
\end{array}
 \]
It is easy to see that knowledge of either $y_1$ or $y_2$ does not
reduce the uncertainty about $s_1$, whereas the knowledge of both
$y_1$ and $y_2$ is sufficient to completely determine $s_1$, namely,
$s_1=y_1+y_2$.

In general, $k=n-\mu$ symbols can be transmitted securely by a
coding scheme based on an $[n,n-k]$ linear maximal distance separable (MDS) code ${\cal
C}\subset\mathbb{F}_q^n$. In this scheme, the encoder is a
probabilistic device which operates on the space $\mathbb{F}_q^n$ partitioned into $q^k$
cosets of ${\cal C}$, where $q$ is a large enough prime power. The $k$ information symbols are taken as the
syndrome which specifies a coset, and the transmitted word is chosen
uniformly at random from the specified coset. The decoder recovers
the information symbols by simply computing the syndrome of the
received word. Because of the properties of MDS codes, knowledge of any
 $\mu=n-k$ or fewer
symbols will leave the uncertainty of the $k$ information symbols
unchanged. The code used in the above example is the $[2,1]$
repetition code with the parity check matrix
\begin{equation}
\mathcal{H}=\begin{bmatrix}
    1 & 1 \\
  \end{bmatrix}.
\label{eq:hex}
\end{equation}

\section{Wiretap Network II\label{sec:wtn}}
We now consider  an acyclic multicast network \mbox{$G=(V,E)$} with
unit capacity edges, an information source, $t$ receivers, and the
value of the min-cut to each receiver is equal to $n$. The goal is to
maximize the multicast rate with the constraint of revealing no
information about the multicast data to the adversary that can
access data on any $\mu$ edges. We assume that the adversary knows
the implemented network code, \emph{i.e.}  all the coefficients of the
linear combinations that determine the packets on each edge.
Moreover, we assume that there  is no shared randomness between the source and the  receivers. The {latter assumption rules out the use of traditional ``key'' cryptography to achieve security.

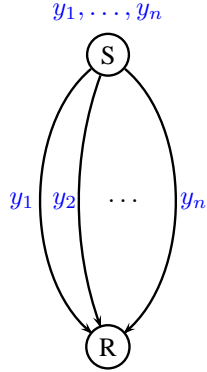
\begin{figure}[hbt]
\begin{center}
\psset{unit=0.045in}

\begin{pspicture}(0,0)(40,45)
\psset{linewidth=0.9pt}

\begin{small}

\rput(20,42){\textcolor{blue}{$y_1,\dots,y_n$}}

\rput(20,37){\circlenode{S}{S}}
\rput(20,3){\circlenode{R}{R}}
\nccurve[ArrowInside=->,ArrowInsidePos=1,
angleA=-140,angleB=140]{S}{R}

\nccurve[ArrowInside=->,ArrowInsidePos=1,
angleA=-110,angleB=110]{S}{R}

\nccurve[ArrowInside=->,ArrowInsidePos=1,
angleA=-40,angleB=40]{S}{R}
\rput(22,20){{$\dots$}}
\rput(10,20){\textcolor{blue}{$y_1$}}
\rput(15,20){\textcolor{blue}{$y_2$}}
\rput(30,20){\textcolor{blue}{$y_n$}}

 \end{small}
\end{pspicture}
\end{center} 
\caption{\label{fig:CNet} Network equivalent to the wiretap channel of type II.}
\end{figure}

 It can be seen that the wiretap channel of type II is equivalent to the simple unicast network of Figure~\ref{fig:CNet} formed by $n$ disjoint edges between the source and the destination, each carrying a different symbol.  For this network, the source
can multicast $k\le n-\mu$ symbols securely if it first applies a  secure wiretap channel code (as described above) mapping $k$ information symbols into $n$ transmitted symbols $(y_1,\dots, y_n)$.

For general networks, when security is not an issue,  we know that a multicast rate  $n$ is possible with linear network coding \cite{Ahl, Li}.  It is interesting to ask whether, using the same network code, the source can always multicast $k\le n-\mu$ symbols securely using a wiretap channel code at the source. Naturally, this would be a solution if a multicast rate
of $n$ can be achieved just by routing.
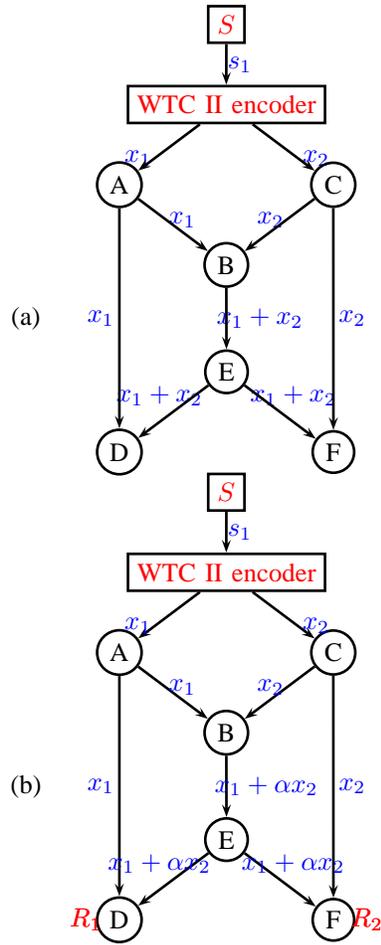
\begin{figure}[hbt]
\begin{center}
\psset{unit=0.070in}
\begin{pspicture}(0,5)(30,78)
\psset{linewidth=1.0pt}
\begin{small}
\rput(16,39){\rnode{S}{\psframebox{\textcolor{red}{$S$}}}}
\rput(16,33){\rnode{SE}{\psframebox{\textcolor{red}{WTC II
encoder}}}} \rput(8,27){\circlenode{A}{A}}
\rput(16,21){\circlenode{B}{B}} \rput(24,27){\circlenode{C}{C}}
\rput(8,7){\circlenode{D}{D}} \rput(16,13){\circlenode{E}{E}}
\rput(24,7){\circlenode{F}{F}} \rput(5.5,7){\textcolor{red}{$R_1$}}
\rput(26.5,7){\textcolor{red}{$R_2$}}
\ncline{->}{S}{SE} \ncline{->}{SE}{A} \ncline{->}{SE}{C}
\ncline{->}{A}{B}\ncline{->}{C}{B}
\ncline{->}{A}{D}\ncline{->}{C}{F}
\ncline{->}{E}{D}\ncline{->}{E}{F} \ncline{->}{B}{E}
\rput(17,71){\textcolor{blue}{$s_1$}}
\rput(9.4,64){\textcolor{blue}{$x_1$}}
\rput(22.7,64){\textcolor{blue}{$x_2$}}
\rput(12.7,59.4){\textcolor{blue}{$x_1$}}
\rput(19.3,59.4){\textcolor{blue}{$x_2$}}
\rput(6.6,52){\textcolor{blue}{$x_1$}}
\rput(25.4,52){\textcolor{blue}{$x_2$}}
\rput(18.5,52){\textcolor{blue}{$x_1+x_2$}}
\rput(11,46.2){\textcolor{blue}{$x_1+ x_2$}}
\rput(21,46.2){\textcolor{blue}{$x_1+x_2$}}

\rput(16,74){\rnode{S}{\psframebox{\textcolor{red}{$S$}}}}
\rput(16,68){\rnode{SE}{\psframebox{\textcolor{red}{WTC II
encoder}}}} \rput(8,62){\circlenode{A}{A}}
\rput(16,56){\circlenode{B}{B}} \rput(24,62){\circlenode{C}{C}}
\rput(8,42){\circlenode{D}{D}} \rput(16,48){\circlenode{E}{E}}
\rput(24,42){\circlenode{F}{F}} \rput(5.5,7){\textcolor{red}{$R_1$}}
\rput(26.5,7){\textcolor{red}{$R_2$}}
\ncline{->}{S}{SE} \ncline{->}{SE}{A} \ncline{->}{SE}{C}
\ncline{->}{A}{B}\ncline{->}{C}{B}
\ncline{->}{A}{D}\ncline{->}{C}{F}
\ncline{->}{E}{D}\ncline{->}{E}{F} \ncline{->}{B}{E}
\rput(17,36){\textcolor{blue}{$s_1$}}
\rput(9.4,29){\textcolor{blue}{$x_1$}}
\rput(22.7,29){\textcolor{blue}{$x_2$}}
\rput(12.7,24.4){\textcolor{blue}{$x_1$}}
\rput(19.3,24.4){\textcolor{blue}{$x_2$}}
\rput(6.6,17){\textcolor{blue}{$x_1$}}
\rput(25.4,17){\textcolor{blue}{$x_2$}}
\rput(19,17){\textcolor{blue}{$x_1+\alpha x_2$}}
\rput(11,11.2){\textcolor{blue}{$x_1+ \alpha x_2$}}
\rput(21,11.2){\textcolor{blue}{$x_1+\alpha x_2$}}
\rput(1,52){(a)}
\rput(1,17){(b)}
\end{small}
\end{pspicture}
\end{center}
\caption{Single-edge wiretap butterfly network
with a) insecure network code and b) secure network code.} \label{fig:butterfly_sec}
\end{figure}
\begin{example}[Butterfly Network]
Consider this approach for the butterfly network
shown in Figure~\ref{fig:butterfly_sec}
where we have $n=2$, $k=1$, $\mu=1$. If the source applies the
coding scheme described in the previous section and the usual
network code as in  Figure~\ref{fig:butterfly_sec}(a), the wiretapper
will be able  to learn the source symbol if  it taps into any
of the edges BE, EF or ED. Therefore, a network code can  break
down a secure wiretap channel code. However, if the network code is
changed so that node B combines its inputs over, {\it e.g.,}
$\mathbb{F}_3$ and the  coding vector of edge BE  is $\begin{bmatrix}
    1 & \alpha \\
  \end{bmatrix}
$ where $\alpha$ is a primitive element of $\mathbb{F}_3$ (\emph{i.e.}, the message sent on edge BE is $x_1+\alpha x_2$ as in Figure~\ref{fig:butterfly_sec}(b)),
the wiretap channel code remains secure, that is, the adversary cannot gain any information by accessing
any single  edge in the network. Note that the wiretap channel code based on the MDS code with
$\mathcal{H}=\begin{bmatrix}
    1 & 1 \\
  \end{bmatrix}$
remains secure with any network code whose BE coding vector is linearly
independent of $\begin{bmatrix}
    1 & 1 \\
  \end{bmatrix}
  $.
\end{example}

We will next show that the source can multicast $k\le n-\mu$ symbols
securely if it first applies a secure wiretap channel code based on
an MDS code with a $k\times n$ parity check matrix $\mathcal{H}$ if the
network code is such that no linear combination of $\mu=n-k$ or
fewer coding vectors belongs to the space spanned by the rows of
$\mathcal{H}$. Let $W\subset E$ denote the set of $|W|=\mu$ edges the
wiretapper chooses to observe, and $Z_W=(z_1,z_2,\dots,z_{\mu})^T$ the
random variable associated with the packets carried by the edges in
$W$. Let $C_W$ denote the matrix whose rows are the coding vectors
associated with the observed edges in $W$. As in the case of the  wiretap
channel, $S=(s_1,s_2,\dots,s_k)^T$ denotes the random variable
associated with the $k$ information symbols that the source wishes
to send securely, and $Y=(y_1,y_2,\dots,y_n)^T$ the random variable
associated with the $n$ wiretap channel code symbols. The $n$ symbols of $Y$ will be multicast through the network by using linear network coding. Writing $H(S,Y,Z_W)$ in two different forms, and taking into account the decodability condition of Equation~\eqref{eq:deccon}, we get
\begin{equation}\label{eq:entropy}
H(S|Z_W)+H(Y|SZ_W)=H(Y|Z_W)+\underbrace{H(S|YZ_W)}_{=0}.
\end{equation}
Our objective  is to conceal all the information data from the wiretapper. The perfect secrecy condition implies
\begin{equation*}
    H(S|Z_W)=H(S),  \forall W\subset E \text{  s.t. } |W|=\mu.
\end{equation*}
Thus we obtain,

\begin{align}
H(Y|SZ_W) &= H(Y|Z_W) - H(S).
% alex - changed \text{rank}(\mathbf{C}_W) to \text{rank}({C}_W)
\end{align}

This implies, in turn that
\begin{align}
n-\text{rank}({C}_W)-k\ge 0.
\end{align}

Since there is a choice of edges such that $\text{rank}(C_W)=\mu$, the maximum rate
for secure transmission is bounded as
\[
k\le n-\mu.
\]
If the bound is achieved with equality, we have \mbox{$H(Y|SZ_W)=0$} and consequently, the system of equations
\[
\begin{bmatrix}
  S \\
  Z_w \\
\end{bmatrix}
= \begin{bmatrix}
    \mathcal{H} \\
    C_W \\
  \end{bmatrix}
  \cdot Y
\]
has to have  a unique solution for all $W$ for which $\text{rank}(C_W)=\mu$. That is,
\begin{equation}
\text{rank}
\begin{bmatrix}
    \mathcal{H}\\
    C_W \\
  \end{bmatrix}
  = n ~~ \text{for all
  $C_W$ s.t.} ~ \text{rank}(C_W)=\mu.
\label{eq:secc1}
\end{equation}
This analysis  proves the following result:
\begin{theorem}
\label{th:oursec}  Let $G=(V,E)$ be an acyclic multicast network with
unit capacity edges and an information source such that the size of a minimum cut between the source and each receiver is equal to $n$.
%
%the  min-cut value to each receiver equal to $n$.
%
Then, a wiretap code at the source based on an
MDS code with a $k\times n$ parity check matrix $\mathcal{H}$ and a network
code such that no linear combination of $\mu=n-k$ or fewer coding
vectors belongs to the space spanned by the rows of $\mathcal{H}$ make the
network  information-theoretically secure against a wiretap adversary
who can observe at most $\mu \le n-k$ edges. Any adversary able to observe
more than $n-k$ edges will have uncertainty about the source smaller
than $k$.
\end{theorem}
Next, we give an application of the previous theorem to the family of \emph{combination} networks illustrated in Figure~\ref{fig:sn}.

%Consider the class  of networks, known as combination networks, that are
\begin{figure}[hbt]
\begin{center}
\psset{unit=0.045in}
\begin{pspicture}(0,0)(70,35)
\psset{linewidth=0.9pt}
\begin{small}
%\rput(35,32){\rnode{S}{\psframebox{\textcolor{red}{$S_1,S_2,\dots,S_h$}}}}
\cnode(35,32){1}{S}
\cnode[linecolor=white](28,38){1}{s1}
%\cnode[linecolor=white](32,38){1}{s2}\ncline{->}{s2}{S}
\cnode[linecolor=white](42,38){1}{sh}
%\rput(29,39){{$S_1$}}
%\rput(42,39){{$S_h$}}
%\rput(35,35){$\cdots$}
%\ncline{->}{s1}{S}\ncline{->}{sh}{S}
\rput(35,29.5){$\cdots$}\rput(35,20){$\cdots$}\rput(35,5){$\cdots$}
\cnode(15,20){1}{u1}
\cnode(20,20){1}{u2}
\cnode[linecolor=white](30,20){1}{u3}
\cnode[linecolor=white](40,20){1}{u3r}
\rput(17,12){$\cdots$}\rput(53,12){$\cdots$}
\psarc[linecolor=blue]{->}(10,5){4}{10}{100}
\psarc[linecolor=blue]{<-}(60,5){4}{80}{170}
\psarc[linecolor=blue]{->}(35,32){4}{190}{350}
\rput(8.2,9){{$n$}}\rput(40,32.5){{$M$}}
\rput(61.8,9){{$n$}}
\ncline{->}{S}{u1}\ncline{->}{S}{u2}
\cnode(50,20){1}{u1r}
\cnode(55,20){1}{u2r}
\ncline{->}{S}{u1r}\ncline{->}{S}{u2r}
\cnode(10,5){1}{r1}
\ncline{->}{u1}{r1}\ncline{->}{u2}{r1}
\ncline{->}{u3}{r1}
\cnode(60,5){1}{r1r}
\ncline{->}{u1r}{r1r}\ncline{->}{u2r}{r1r}
\ncline{->}{u3r}{r1r}
\rput(10,2){{$R_1$}}
\rput(60,2){{$R_{\binom{M}{n}}$}} \end{small}
\end{pspicture}
\end{center} 
\caption{\label{fig:sn}Combination $B(n,M)$ network.}
\end{figure}
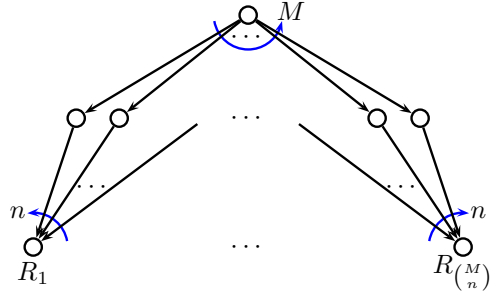
\begin{example}[Combination Networks]
A combination network $B(n,M)$ is defined over a 3-partite graph comprising  three layers. The first layer contains a single source node, the second layer $M$ intermediate nodes and the last  layer is formed by  $\binom{M}{n}$ receiver nodes  such that every set of  $n$ nodes of the second layer is observed by a receiver.

  The result of  Theorem~\ref{th:oursec} can be used to construct a secure network code for $B(n,M)$  from an  $[M+k,M+k-n]$ MDS code which would achieve perfect secrecy against a wiretapper that can observe any $\mu=n-k$  edges in the network. Let $\mathcal{H}$ be an $n\times (M+k)$ parity check matrix of such MDS code over $\mathbb{F}_q$. A secure network code can be obtained by taking  the first $k$ rows of $\mathcal{H}^T$ to form the matrix of the coset code at the source, and the rest of the rows of $\mathcal{H}^T$ to be the coding vectors of the $M$ edges going out of the source. Equation \eqref{eq:secc1} is satisfied since the considered code is MDS and, therefore, any $n$ columns of $\mathcal{H}$ form a basis of $\mathbb{F}_q^n$. For instance if $M+k+1$ is equal to a  prime power $q$, a secure network code can be derived based on an $[M+k,M+k-n]$ Reed-Solomon code with  the following Vandermonde parity check matrix
\begin{equation}
\label{eq:RSParity}
\mathcal{H}=
\begin{bmatrix}
    1&\alpha&\dots&\alpha^{M+k-1} \\
    1&\alpha^2&\dots&\alpha^{2(M+k-1)} \\
    \vdots&\vdots&\vdots&\vdots\\
    1&\alpha^{n}&\dots&\alpha^{n(M+k-1)}
  \end{bmatrix},
 \end{equation}
  where $\alpha$ is a primitive element of $\mathbb{F}_q$. Figure \ref{fig:RS} depicts a secure network code for the network $B(3,4)$ and $k=2$ using a [6,3] Reed-Solomon code over $\mathbb{F}_7$ whose parity check  matrix is given by  Equation~\eqref{eq:RSParity} for $\alpha=3$.
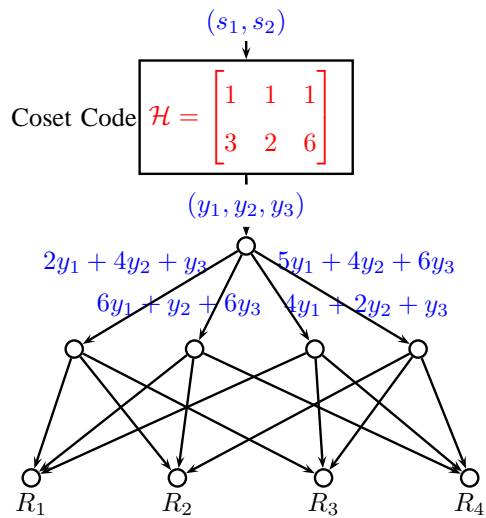
\begin{figure}[hbt]
\begin{center}
\psset{unit=0.045in}
\begin{pspicture}(0,0)(70,60)
\psset{linewidth=0.9pt}
\begin{small}
%\rput(35,32){\rnode{S}{\psframebox{\textcolor{red}{$S_1,S_2,\dots,S_h$}}}}
%\cnode(35,32){1}{S}
\cnode[linecolor=white](28,38){1}{s1}
%\cnode[linecolor=white](32,38){1}{s2}\ncline{->}{s2}{S}
\cnode[linecolor=white](42,38){1}{sh}
%\rput(29,39){{$S_1$}}
%\rput(42,39){{$S_h$}}
%\rput(35,35){$\cdots$}
%\ncline{->}{s1}{S}\ncline{->}{sh}{S}

%\cnode[linecolor=white](35,38){1}{v}
\cnode[linecolor=white](35,57){1}{v}

\rput(35,58){{\textcolor{blue}{$(s_1,s_2)$}}}
\rput(35,47){\rnode{SE}{\psframebox{\textcolor{red}{
$\mathcal{H}=\begin{bmatrix}
1 & 1&1\cr
3 & 2&6
\end{bmatrix}$
}}}}

\rput(15,47){Coset Code}
\ncline{->}{v}{SE}

\cnode(35,32){1}{S}
%\rput(35,36){{\textcolor{blue}{$(y_1,y_2,y_3)$}}}

\ncline{->}{SE}{S}

\ncline[nodesep=1pt]{SE}{S}
\mput*{{\textcolor{blue}{$(y_1,y_2,y_3)$}}}

%\rput(35,29.5){$\cdots$}\rput(35,20){$\cdots$}\rput(35,5){$\cdots$}
\cnode(15,20){1}{u1}
\cnode(29,20){1}{u2}
\cnode(43,20){1}{u3}
\cnode(55,20){1}{u4}

\cnode(10,5){1}{r1}
\cnode(27,5){1}{r2}
\cnode(44,5){1}{r3}
\cnode(61,5){1}{r4}

\rput(21,30){\textcolor{blue}{$2y_1+4y_2+y_3$}}
%\rput(22,25){\textcolor{blue}{$6y_1+y_2+6y_3$}}
\rput(27.25,25.1){\textcolor{blue}{$6y_1+y_2+6y_3$}}
\rput(49,25){\textcolor{blue}{$4y_1+2y_2+y_3$}}
\rput(49,30){\textcolor{blue}{$5y_1+4y_2+6y_3$}}

\ncline{->}{S}{u1}
\ncline{->}{S}{u2}
\ncline{->}{S}{u3}
\ncline{->}{S}{u4}

\ncline{->}{u1}{r1}\ncline{->}{u2}{r1}\ncline{->}{u3}{r1}

\ncline{->}{u1}{r2}\ncline{->}{u2}{r2}\ncline{->}{u4}{r2}

\ncline{->}{u1}{r3}\ncline{->}{u3}{r3}\ncline{->}{u4}{r3}

\ncline{->}{u2}{r4}\ncline{->}{u3}{r4}\ncline{->}{u4}{r4}

\ncline{->}{u1r}{r1r}\ncline{->}{u2r}{r1r}
\ncline{->}{u3r}{r1r}
\rput(10,2){{$R_1$}}
\rput(61,2){{$R_4$}}
\rput(27,2){{$R_2$}}
\rput(44,2){{$R_3$}}
 \end{small}
\end{pspicture}
\end{center} 
\caption{\label{fig:RS} A secure network code for the  $B(3,4)$ combination network based on a [6,3] Reed-Solomon code over $\mathbb{F}_7$.}
\end{figure}
\end{example}

The above analysis shows that the maximum throughput can be achieved
by applying a wiretap channel code at the source and then designing
the network code while respecting certain constraints. The decoding
of secure source symbols $S$ is then merely a matrix multiplication of
the decoded multicast symbols $Y$ since $\mathcal{H}Y=S$. The
method gives us a better insight of how much information the
adversary gets if he can access more edges than the code is designed
for.
 It also enables us to design secure network coding schemes over smaller alphabets.
%It also gives us an insight on how to simply design secure network codes in some cases over much smaller alphabets then currently deemed necessary.
These two issues are discussed in  detail in the next two sections.

\section{Network Code Design Alphabet Size\label{sec:cd}}
The approach described previously in the literature for finding
a secure multicast network code consisted of decoupling the problem
of designing a multicast network code and making it secure by using
some code on top of it.
Feldman \emph{et al.}\ showed in \cite{feldman04csnc}
that there exist networks where the above construction might require
a quite large field size.
 In this section, we present a different construction that exploits the topology of the network.
%We investigate here a different
%construction that, as was hinted in the conclusion of \cite{feldman04csnc}, exploits the topology of the network.
%alex - I have removed as was hinted in the conclusion of \cite{feldman04csnc}
This is accomplished  by adding the security constraints  to the
\emph{Linear Information Flow} (LIF) algorithm of \cite{jaggi03polynomial} that constructs linear multicast network codes in polynomial time in the number of edges in the graph. The result is a better lower bound on the sufficient field size. However, the modified LIF algorithm does not have a polynomial time complexity.

We start by giving a brief high level overview of the LIF algorithm of
\cite{jaggi03polynomial}. The inputs of the algorithm are the network, the source node,
the $t$  receivers and the number $n$ of packets that need to be multicast to all
 receivers. Assuming the min-cut between the source and any
 receiver is at least $n$, the algorithm outputs a linear network
code that guaranties the delivery of the $n$ packets to all the receivers.

The algorithm starts by 1) finding $t$ flows $F_1, F_2,\dots, F_t$
of value $n$ each, from the source to to each  receiver and 2)
 defining $t$ $n \times n$ matrices $B_{F_j}$ (one for each receiver)  formed by the global encoding vectors of the $n$ last visited edges in the flow $F_j$. Initially, each matrix $B_{F_j}$ is equal to the identity matrix $I_{n}$.  Then,  the algorithm goes over the network edges, visiting each one in   a topological order.  In each iteration, the algorithm finds a suitable local encoding vector for the visited
edge, and updates  all of the  $t$ matrices $B_{F_j}$. The algorithm maintains the invariant that the matrices
$B_{F_j}$ remain invertible after each iteration. Thus, when it
terminates, each  receiver will get $n$ linear  combinations of the original packets that form a full rank system. Thus each destination
can solve for these packets by inverting the corresponding  matrix.

 The analysis of the algorithm due to \cite{jaggi03polynomial} implies that a field of size at least  $t$ (the number of destinations) is   sufficient for finding the desired network code.  In particular, as shown in \cite[Lemma 8]{jaggi03polynomial},  a field of
%this follows from the fact that
size larger or equal to $t$ is  sufficient for satisfying
the condition that the $t$ matrices $B_{F_j}$ are always
invertible.

 To construct a secure network code, we modify the LIF algorithm in the following way.
%We modify the LIF algorithm so it  outputs a secure network code in
%the following way.
We  select a $k\times n$ parity check matrix $\mathcal{H}$.  Without loss of generality, we assume that the $\mu$ packets observed
by the wiretapper are linearly independent, \emph{i.e.},\ rank $C_W=\mu$.
We denote by $e_i$ the edge visited at the $i$-th
iteration of the LIF algorithm, and by $P_i$ the set of the edges
that have been processed by the end of it. Then, we extend the set
of invariants to guaranty that the encoding vectors are chosen so
that the  matrices  $M_W=\chsfk{\mathcal{H}}{C_W}$ are also invertible; which, by Theorem~\ref{th:oursec}, achieves the security condition. More
precisely, using the same techniques as the original LIF algorithm,
we make sure that by the end of the $i$-th iteration, the matrices
$B_{F_j}$ and the matrices $M_{W_i}$ are invertible; where
$W_i=\{e_i\}\cup W'$ and $W'$ is a subset of  $P_i$ containing
$\mu-1=n-k-1$ edges. The total number of  matrices that need to be kept
invertible in this modified version of the LIF algorithm is at most
$\binom{|E|-1}{\mu-1}+t$.
%alex - removed - I not understand this sentence (which corresponds to the last iteration).
Thus, similarly as in \cite[Lemma 8]{jaggi03polynomial}, we obtain
the following improved bound on the alphabet size for secure
multicast:
\begin{theorem}
Let $G=(V,E)$ be an acyclic network with
unit capacity edges  and an information source such that the  min-cut value to each of the $t$ receivers  is equal to $n$.
A secure multicast at rate $k\le n-\mu$ in the presence of a wiretapper
who can observe at most $\mu\le n$ edges
is possible over the alphabet $\mathbb{F}_q$ of size
\begin{equation}
q\geq\binom{|E|-1}{\mu-1}+t.
\label{eq:ourfs}
\end{equation}
\label{thbetterbound}
\end{theorem}

The bound given by Equation (\ref{eq:ourfs}) can be further improved by realizing as was first done in \cite{ceitnc} that not all edges in the network carry
different linear combination of the source symbols.  Langberg \emph{et al.}\ showed in \cite{lang06com} that the number of \emph{encoding edges} in a \emph{minimal} acyclic multicast network is bounded by  $2n^3t^2$. Encoding edges create new packets by combining the packets received over the incoming edges of their tail nodes.
% a linear combination of the packets received by their tail nodes.
A minimal multicast network does not contain redundant edges, \emph{i.e.}, edges that can be removed from the network without violating its optimality. Reference \cite{lang06} presents an efficient algorithm for construction of a minimal acyclic network $\widehat{G}$ from the original network $G$. This work also shows that a feasible network code for a minimal network can be used for the original network as well with only slight modifications.

%alex -removed More specifically, \cite{lang06com} shows that the number of encoding edges in a minimal network is bounded by $2n^3t^2$.

%
%
%the problem of finding
%multicast network codes for a network $G$ can be reduced to solving
%the same problem for a special equivalent network $\widehat{G}$ with
%same parameters $n$ and $t$, which has the properties that all nodes
%except the source and the destinations have total degree 3 and at
%most $n^3t^2$ of its nodes have in-degree 2. These nodes are called
%\emph{encoding nodes}, whereas the other ones are called
%\emph{forwarding nodes} since the packets carried by their outgoing
%edges are just copies of the ones available at their single incoming
%edge. Given a network code for $\widehat{G}$, a one for $G$ can be
%found efficiently over the same field. And, the set of global
%encoding vectors of the edges of $G$ would be a subset of the one of
%$\widehat{G}$.

%Going back the security problem over a network $G$, one can try to find a

 The main idea of our scheme is to find a
secure network code for the minimal network $\widehat{G}$,
and then use the procedure described in \cite{lang06} to construct a network code for  original network $G$ which will also be secure. Now consider the problem of finding secure network codes for $\widehat{G}$. This problem will not change if the wiretapper is not allowed to wiretap the \emph{forwarding edges}, \emph{i.e.}, the edges that just forward packets received by their tail nodes.  Therefore, the
set of edges that the wiretapper might have access to consists of the encoding edges and the edges outgoing from the source.  The number of such edges is bounded by  $2n^3t^2$. Now, applying Theorem \ref{thbetterbound} on $\widehat{G}$ and taking into consideration the restriction on the edges that can
be potentially wiretapped, we obtain the following bound on the sufficient field size which is independent of the size of the
network.
\begin{corollary}
For the transmission scenario of Theorem~\ref{thbetterbound}, a secure mulitcast network code always exists over the alphabet
$\mathbb{F}_q$ of size
\begin{equation}
q\geq\binom{2k^3t^2}{\mu-1}+t. \label{eq:ourifs}
\end{equation}
\end{corollary}
\vspace{2mm}

For networks with two sources, we can completely settle the question on the required alphabet
size for a secure network code. Note that the adversary has to be limited to observing at most one edge of his
choice. Based on the work of Fragouli and Soljanin in \cite{ceitnc},
the coding problem for these networks is equivalent to a vertex coloring problem of some specially
designed graphs, where the colors  correspond to the points on
the projective line $\mathbb{PG}(1,q)$:
\begin{equation}
[0\, 1], ~ [1\, 0], ~\text{and} ~ [1 \, \alpha^i] ~ \text{for} ~ 0\le i\le q-2,
\label{eq:plp}
\end{equation}
where $\alpha$ is a primitive element of  $\mathbb{F}_q$.
Clearly, any network with two sources and arbitrary number of receives can be securely coded by reducing the set of available
colors in (\ref{eq:plp}) by removing point (color) $[1\, 1]$ and applying a wiretap code based on the matrix $\mathcal{H}=[1\, 1]$ as in the
example above.
Alphabet size sufficient to securely code all network with two sources also follows from \cite{ceitnc}:
\begin{theorem}
\label{th:h2st2}
For any configuration with two sources $t$ receivers,
the code alphabet $\mathbb{F}_q$ of size
\[
\lfloor \sqrt{2t-7/4}+1/2 \rfloor +1
\label{eq_alph}
\]
is sufficient for a secure network code. There exist configurations for which it is necessary.
\end{theorem}

%\newpage

\section{Wiretapper Equivocation}\label{sec:WiretapperEquivocation}

In this section, we analyze the performance of coset codes in  the case of a   wiretapper with variable strength, \emph{i.e.}, the number $\mu$ of edges he can observe is not fixed. For a given coset code, we seek to quantify the amount of information that is leaked to the wiretapper as a function of $\mu$.

Assume that at the source $s$ of a multicast network a coset code defined by a $k\times n$ parity check matrix $\mathcal{H}$ is used as described in the previous section.  The equivocation $\Delta(\mu)$ of the wiretapper, \emph{i.e.}, the uncertainty it has about the information source vector $S=(s_1,\dots,s_k)^T$, is defined, as in \cite{OzarowWy85}, based on the worst case scenario, by
\begin{equation}\label{eq:equi}
    \Delta(\mu):=\min_{W\subset E; |W|=\mu} H(S|Z_W),
    \end{equation}
where $Z_W=(z_1,\dots,z_\mu)^T$  is the random  variable representing the observed packets on the set  $W\subseteq E$ of wiretapped edges. We have $Z_W=C_W Y$ where $C_W$ is an $\mu\times n$ matrix, and $Y=(y_1,\dots,y_n)^T$ is the output of the coset code at the source. It can be seen that $\Delta(\mu)$ can be written as:

\begin{equation}\label{eq:equi2}
    \Delta(\mu)=\min_{\substack{W\subset E; |W|=\mu\\\rank(C_W)=\mu}} H(S|Z_W).
    \end{equation}

    Therefore, we will assume from now on without loss of generality that  $W$ is such that $\rank(C_W)=\mu$.  For a given choice of such $W$, let $C_W^\perp$ be the parity check matrix of the $[n,\mu]$ code generated by $C_W$. Let $I_n$ be the $n\times n $ identity matrix. Define $J_{n,\mu}$ to be the $n\times (n-\mu)$ matrix where the first $\mu$ rows are all zeros and the last $n-\mu$ rows form $I_{n-\mu}$.     Theorem~\ref{th:equi} below  gives the expression of $\Delta(\mu)$ which depends on the network code  and the coset code used.

    \begin{theorem}\label{th:equi}
    \begin{equation}\label{eq:th2}
    \Delta(\mu)=\min_{\substack{W\subset E; |W|=\mu\\\rank(C_W)=\mu}} \rank(\mathcal{H}\begin{bmatrix}
    C_W \\
    C_W^\perp \\
  \end{bmatrix}^{-1}J_{n,\mu}).
    \end{equation}
    \end{theorem}

    \begin{proof}

     First let
     $A_W=\begin{bmatrix}
    C_W \\
    C_W^\perp \\
  \end{bmatrix}.$ By  Equation~\eqref{eq:entropy}, we have

 \begin{equation}
\begin{split}
H(S|Z_W)& = H(Y|Z_W)-H(Y|SZ_W)\\
& = n-\rank(C_W)-(n-\rank\begin{bmatrix}    \mathcal{H} \\    C_W\\  \end{bmatrix})\\
& = \rank(\begin{bmatrix}   \mathcal{H} \\    C_W\\  \end{bmatrix}A_W^{-1})-\rank(C_W)\\
& = \rank(\begin{bmatrix}   \mathcal{H}A_W^{-1} \\    C_WA_W^{-1}\\  \end{bmatrix})-\rank(C_W)\\
&= \dim(\langle \mathcal{H}A_W^{-1}\rangle)+\dim(\langle C_WA_W^{-1}\rangle)\\
&-\dim(\langle \mathcal{H}A_W^{-1}\rangle\cap\langle C_WA_W^{-1}\rangle)-\rank(C_W)\\
&=k-\dim(\langle \mathcal{H}A_W^{-1}\rangle\cap \langle J'_{n,\mu}\rangle),
\end{split}
\end{equation}
where $\langle\cdot\rangle$ denotes the row space of a matrix and   $J'_{n,\mu}$ is the $\mu\times n$ matrix where the first $\mu$ columns form  $I_\mu$   and the last $n-\mu$ columns are all zeros. Note that \mbox{$\dim(\langle \mathcal{H}A_W^{-1}\rangle\cap \langle J'_{n,\mu}\rangle)$} is exactly  $k$  minus the rank of the last $n-\mu$ column vectors of $\mathcal{H}A_W^{-1}$.
    \end{proof}

 A relevant concept to our work here is that of the generalized Hamming weights  $d_1(\mathcal{C}),\dots,d_{k}(\mathcal{C})$ of a linear code $\mathcal{C}$ which was introduced by Wei in \cite{Wei91} and that characterize the performance of coset codes over the classical wiretap channel of type II.   The generalized Hamming weights were  extended to the
%alex - you had generalized twice
wiretap networks setting in \cite{CYZ09}. Given a  certain network with an associated network and coset codes, Theorem \ref{th:equi} provides an equivalent expression of the network formulation of the $r$-th generalized Hamming weight $d_r$ as the minimum number of edges that should be wiretapped to leak $r$ symbols to the wiretapper. Then, we  can write

 \begin{equation}
\begin{split}
d_r &:= \min \{\mu; \Delta(\mu)=k-r\}\\
& := \min \{\mu; \min_{\substack{W\subset E; |W|=\mu\\\rank(C_W)=\mu}}\rank(\mathcal{H}\begin{bmatrix}
    C_W \\
    C_W^\perp \\
  \end{bmatrix}^{-1}J_{n,\mu})=k-r\}.
\end{split}
\end{equation}

    Next, we focus on three special cases. First, we revisit the model of the  wiretap channel of type II of \cite{Ozarow&Wyner:84}. Second, we consider the case where the wiretapper may gain access to more edges  than what the secure code is designed to combat. Third, we study the scenario where only a part of the network edges are vulnerable to wiretapping.
   \subsection{Wiretap Channel of Type II}
   Consider again the wiretap channel of type II studied in \cite{Ozarow&Wyner:84}. Theorem \ref{th:equi} can be used to easily recover the following classical result for this channel.

    \begin{corollary}\label{cor:WTCII}
    The equivocation rate of the wiretapper in the wiretap channel of type II is given by
   \begin{equation}\label{eq:OZequi}
    \Delta(\mu)=\min_{\substack{U\subseteq\{1,2,\dots,n\}\\|U|=n-\mu}}\rank \{ \mathcal{H}_i; i\in U\},
    \end{equation}
    where $\mathcal{H}_i$ denote the ith column of the parity check matrix $\mathcal{H}$.
    \end{corollary}
    \begin{proof}
    The wiretap channel of type II is equivalent to the network depicted in  Figure~\ref{fig:CNet}. Assume that the edges between the source and the destination are indexed from 1 to $n$, so that $E=\{1,\dots,n\}$. For any $W\subseteq \{1,\dots,n\}$, define $I_W$ to be the matrix formed by the rows of the $n\times n$ identity matrix indexed by the elements of $W$ in an increasing order. Since edge $i$ carries the packet $y_i$, for a given set $W\subseteq E$ of wiretapped edges, $C_W=I_W$ and $C_W^\perp=I_U$, where $U=\{1,\dots,n\}\setminus W$. Therefore, $A_W^{-1}=\begin{bmatrix}
    I_W \\
    I_U \\
  \end{bmatrix}^{-1}=A_W^T,$ and the last $n-\mu$ columns of $\mathcal{H}A_W^T$ are exactly the columns of $\mathcal{H}$  indexed by $U$.
    \end{proof}

    \subsection{Underestimated Wiretapper}
    Suppose the coset code defined by the $k\times n$ parity check matrix $\mathcal{H}$ satisfies Theorem~\ref{th:oursec} and achieves perfect secrecy against a wiretapper that can observe $\lambda$ edges. If, however, the wiretapper can access $\mu$ edges, where  $\mu>\lambda$, then the amount of information  leaked to the wiretapper can be shown to be   equal to $\mu-\lambda$, \emph{i.e.}, the number of additional wiretapped edges.
    \begin{corollary} For the case of an underestimated wiretapper, the equivocation of the wiretapper is given by: $$\Delta(\mu)=k-(\mu-\lambda).$$
    \end{corollary}
    \begin{proof}
    Since the coset code achieves perfect secrecy for $\lambda$ wiretapped edges, by Theorem~\ref{th:oursec}, we have $k=n-\lambda$ and $H(S|YZ_W)=0$. Thus, Equation~\eqref{eq:entropy} gives
    $$H(S|Z_W) = H(Y|Z_W)= n-\rank(C_W) =  k+\lambda-\rank(C_W).$$ The minimum value of $H(S|Z_W)$ is obtained when $C_W$ has maximal rank, i.e, when \mbox{$\rank(C_W)=\mu$.}  \end{proof}

    \subsection{Restricted Wiretapper}
    In practice, for instance in large  networks, the wiretapper may not have access to all the network edges, and his choice of $\mu$ edges is limited to a certain edge subset $E'\subset E$. For this model, the equivocation rate of the wiretapper is determined by Equation~\ref{eq:th2} where $E$ is replaced by $E'$. An interesting case arises, however, when the edges in $E'$ belong to a cut of $n$ edges between the source and one of the receivers. In this case, the performance of the coset code is the same  as when it is used for a wiretap channel of type II.

    \begin{corollary} In the case of a restricted wiretapper that can  observe any $\mu$ edges in a cut between the source and one of the destinations, the equivocation rate of the wiretapper is given by Equation~\eqref{eq:OZequi}.
     \end{corollary}
    \begin{proof}
    Assume the edges that  are vulnerable to wiretapping are indexed from 1 to $n$, so that $E'=\{1,\dots,n\}$. Let $Z_{E'}=(z_1,\dots,z_n)^T$ denote the packets carried by those edges, such that edge $i$ carries packet $z_i$. We can write $Z_{E'}=C_{E'}Y$, where $C_{E'}$ is an  $n\times n$ matrix. Since the cut comprises $n$ edges, the matrix $C_{E'}$ is invertible; otherwise, by the properties of linear network codes, the destination corresponding to the considered cut cannot decode $Y$. For a choice $W\subseteq E'$ of wiretapped edges, we have $Z_W=C_W Y$, where $C_W=I_WC_{E'}$. Moreover, $C_W^\perp=I_{\overline{W}}C_{E'}$, where $\overline{W}=E'\setminus W$.
    Therefore,
    $$\mathcal{H}\begin{bmatrix}
    C_W \\
    C_W^\perp \\
  \end{bmatrix}^{-1}=\mathcal{H}(C_{E'} \begin{bmatrix}
    I_W \\
    I_{\overline{W}} \\
  \end{bmatrix})^{-1}=\mathcal{H}C_{E'}^{-1}\begin{bmatrix}
    I_W \\
    I_{\overline{W}} \\
  \end{bmatrix}^T.$$
  Similar to the proof of   Corollary~\ref{cor:WTCII}, the last $n-\mu$ columns of $\mathcal{H}A^{-1}\begin{bmatrix}
    I_W \\
    I_{\overline{W}} \\
  \end{bmatrix}^T$ are exactly the columns of $\mathcal{H}A^{-1}$  indexed by $U$. So, by  Theorem~\ref{th:equi}, we have

  \begin{equation*}
\begin{split}
  \Delta(\mu)&=\min_{\substack{U\subseteq\{1,2,\dots,n\}\\|U|=n-\mu}}\rank \{(\mathcal{H}A^{-1})_i;i\in U\}\\
 &=\min_{\substack{U\subseteq\{1,2,\dots,n\}\\|U|=n-\mu}}\rank \{ \mathcal{H}_i; i\in U\}.\\
\end{split}
\end{equation*}
\end{proof}

Note that the previous result  still holds for any subset $E'$ of possible wiretapped edges such that $C_{E'}$ is invertible. For this scenario, the equivocation rate of the wiretapper can be alternatively given by the generalized Hamming weights \cite{Wei91} $d_1(\mathcal{C}),\dots,d_{k}(\mathcal{C})$ of the linear code $\mathcal{C}$   generated by $\mathcal{H}$. In this case, for a given $\mu$, $\Delta(\mu)$ is the unique solution to the following inequalities \cite[Cor. A]{Wei91}:

$$d_{n-\mu-\Delta(\mu)}(\mathcal{C}))\leq n-\mu<d_{n-\mu-\Delta(\mu)+1}(\mathcal{C}).$$

\begin{figure*}[t]
\begin{center}
	\begin{center}
\psset{unit=0.045in}
\begin{pspicture}(0,0)(110,20)
\psset{linewidth=0.9pt}
\begin{small}
%\rput(35,32){\rnode{S}{\psframebox{\textcolor{red}{$S_1,S_2,\dots,S_h$}}}}
%\cnode(35,32){1}{S}
%\cnode[linecolor=white](28,38){1}{s1}
%\cnode[linecolor=white](32,38){1}{s2}\ncline{->}{s2}{S}
%\cnode[linecolor=white](42,38){1}{sh}
%\rput(29,39){{$S_1$}}
%\rput(42,39){{$S_h$}}
%\rput(35,35){$\cdots$}
%\ncline{->}{s1}{S}\ncline{->}{sh}{S}

%\cnode[linecolor=white](35,38){1}{v}

\cnode[linecolor=white](10,10){1}{v}
\rput(10,13){{\textcolor{blue}{$(s_1,\dots,s_k)$}}}
\rput(25,10){\rnode{SE}{\psframebox{\textcolor{red}{
$\quad H \quad$
}}}}
\rput(25,5){{\textcolor{black}{coset code}}}

%\rput(10,18){{\textcolor{blue}{$(s_1,\dots,s_k)$}}}
\rput(55,10){\rnode{NE}{\psframebox{\textcolor{red}{
$\quad G^\perp \quad$
}}}}
\ncline{->}{SE}{NE}

\ncline{->}{v}{SE}
\rput(39,13){{\textcolor{blue}{$(t_1,\dots,t_{m})$}}}
\rput(55,5){{\textcolor{black}{Network}}}
\rput(55,2.4){{\textcolor{black}{Error-Correcting}}}
\rput(55,.3){{\textcolor{black}{Code}}}

\rput(85,10){\rnode{NT}{\psframebox{\textcolor{red}{
Network
}}}}
\ncline{->}{NE}{NT}
\rput(70,13){{\textcolor{blue}{$(y_1,\dots,y_{n})$}}}

\ncline{->}{t}{NT}

\cnode[linecolor=white](100,10){1}{vv}
\ncline{->}{NT}{vv}
%\rput(35,36){{\textcolor{blue}{$(y_1,y_2,y_3)$}}}

%\ncline{->}{SE}{S}

%\ncline[nodesep=1pt]{SE}{S}
%\mput*{{\textcolor{blue}{$(y_1,y_2,y_3)$}}}

%%\rput(35,29.5){$\cdots$}\rput(35,20){$\cdots$}\rput(35,5){$\cdots$}
%\cnode(15,20){1}{u1}
%\cnode(29,20){1}{u2}
%\cnode(43,20){1}{u3}
%\cnode(55,20){1}{u4}
%

%\cnode(10,5){1}{r1}
%\cnode(27,5){1}{r2}
%\cnode(44,5){1}{r3}
%\cnode(61,5){1}{r4}
%
%
%
%
%\rput(21,30){\textcolor{blue}{$2y_1+4y_2+y_3$}}
%\rput(22,25){\textcolor{blue}{$6y_1+y_2+6y_3$}}
%\rput(49,25){\textcolor{blue}{$4y_1+2y_2+y_3$}}
%\rput(49,30){\textcolor{blue}{$5y_1+4y_2+6y_3$}}
%
%
%
%
%\ncline{->}{S}{u1}
%\ncline{->}{S}{u2}
%\ncline{->}{S}{u3}
%\ncline{->}{S}{u4}
%
%
%
%\ncline{->}{u1}{r1}\ncline{->}{u2}{r1}\ncline{->}{u3}{r1}
%
%\ncline{->}{u1}{r2}\ncline{->}{u2}{r2}\ncline{->}{u4}{r2}
%
%\ncline{->}{u1}{r3}\ncline{->}{u3}{r3}\ncline{->}{u4}{r3}
%
%\ncline{->}{u2}{r4}\ncline{->}{u3}{r4}\ncline{->}{u4}{r4}
%
%\ncline{->}{u1r}{r1r}\ncline{->}{u2r}{r1r}
%\ncline{->}{u3r}{r1r}
%\rput(10,2){{$R_1$}}
%\rput(61,2){{$R_4$}}
%\rput(27,2){{$R_2$}}
%\rput(44,2){{$R_3$}}
 \end{small}
\end{pspicture}
\end{center} 
\caption{A coding scheme achieving perfect secrecy against a limited  Byzantine wiretapper.} \label{fig:Biz}
\end{center}
\end{figure*}
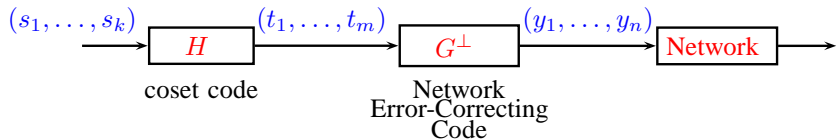

\section{Connections with Other Schemes\label{sec:con}}
 In this section, we explore the relationship  between the proposed scheme and previously known constructions \cite{yeung02secure,zhang06,yy07, SK08}.

\subsection{Secure Network Coding and Filtered Secret Sharing}
Cai and Yeung were first to study the design of secure
network codes for multicast demands \cite{yeung02secure}. They showed
that, in the setting described above, a secure network code can be
found for any $k\leq n-\mu$. Their construction is
equivalent to the following scheme:
\begin{enumerate}
\item Generate a vector $R=(r_1,r_2,\dots,r_{\mu})^T$ choosing its components uniformly
  at random over $\mathbb{F}_q$,
  \item Form vector $X$ by concatenating the $\mu$ random symbols $R$ to the $k$ source symbols $S$:
  \[
  X=\chsfk{S}{R}=(s_1,\dots,s_k,r_1,\dots,r_{\mu})^T
  \]
  \item Chose an \emph{invertible } $n\times n$ matrix  $T$ over $\mathbb{F}_q$ and a  feasible multicast network code \cite{Li}
  to ensure the security condition (\ref{eqsecure}).
  (It is shown in \cite[Thm.~1]{yeung02secure} that such  code and matrix   $T$ can be found provided that $q>\binom{|E|}{\mu}$.)
  \item Compute $Y=TX$ and multicast $Y$ to all the destinations by using the constructed code.
\end{enumerate}

Feldman \emph{et al.}\ considered also the same problem in \cite{feldman04csnc}.
Adopting the same approach of \cite{yeung02secure}, they showed that in order
for the code to be secure, the matrix $T$ should  satisfy certain
conditions (\cite[Thm.~6]{feldman04csnc}).
%alex - deleted that we restate here for convenience:
In particular, they showed that in the above transmission scheme, the security condition (\ref{eqsecure}) holds if and only if
any set of vectors consisting of
\begin{enumerate}
  \item at most  $\mu$ linearly independent  edge coding vectors and/or
  \item any number of vectors from the first $k$ rows of $T^{-1}$
\end{enumerate}
is linearly independent.
They also
showed that if one sacrifices in the number of information
packets, that is, take $k< n-\mu$, then  it is possible to find secure network
codes over fields of size much smaller than the very large bound $q>\binom{|E|}{\mu}$.

We will now show that our approach based on coding for the wiretap channel at the source is equivalent to
the above stated scheme \cite{yeung02secure} with the conditions of \cite{feldman04csnc}.
\begin{proposition}
For any  $n\times n$ matrix $T$
% alex - I would not introduce a new definition - {\cal NC} - we did not define network coding formally
 satisfying the security conditions defined  above, the $k\times n$ matrix  $\mathcal{H}=T^*$  formed by taking the first $k$ rows
of $T^{-1}$ satisfy the condition of
Theorem~\ref{th:oursec}. \label{eqsecure}\end{proposition}
\begin{proof}
Consider the secure multicast scheme of \cite{yeung02secure} as
presented above. For a given information vector $S\in
\mathbb{F}_q^k$, let $B(S)$ be the set of all possible vectors $Y\in
\mathbb{F}_q^{n}$ that could be multicast through the network under
this scheme. More precisely,
\begin{equation*}
    B(S)=\Bigl\{Y\in \mathbb{F}_q^n | Y=TX,  X=\chsfk{S}{R}, R\in\mathbb{F}_q^{n-k}\Bigr\}.
\end{equation*}
Then, for all $Y\in B(S)$, we have $ T^*Y=T^*T\chsfk{S}{T}=S. $
Therefore, any $Y\in B(S)$ also belongs to the coset of the space
spanned by the rows of $T^*$ whose syndrome is equal to $S$.
Moreover, since $T$ is invertible, $|B(S)|=2^{n-k}$ implying that
set $B(S)$ is exactly that coset. The conditions of
\cite{feldman04csnc} as stated above directly translate into
(\ref{eq:secc}), the remaining condition of Theorem~\ref{th:oursec}.
\end{proof}

\subsection{Universal Secure Network Codes}

For practical implementations of linear multicast network codes over $\mathbb{F}_q$, the information sources are typically packets of a certain length $m$, \emph{i.e.}, $s_1,\dots,s_k$ are vectors in $\mathbb{F}_q^m$. Applying the approach presented  in the preliminary version of this paper \cite{RS07}, Silva and Kschischang devised in
% should we say- presented in the preliminary version of this paper \cite{RS07}
\cite{SK08} a scheme  that achieves a complete decoupling between the secure code and the network code design. Their scheme is universal in the sense that it achieves  secrecy by applying a coset code at source
%alex - deleted on top of the network
with no knowledge  of the network code used. The main idea is to use a special
% alex - removed "or change"
class of MDS codes called maximal rank-distance codes (MRD) which are non-linear over $\mathbb{F}_q$ but linear over the extension field $\mathbb{F}_{q^m}$. The parity check matrix of an MRD code over  $\mathbb{F}_{q^m}$,  has the interesting property that it always satisfies the condition of Theorem~\ref{th:oursec} when the edge coding vectors are over $\mathbb{F}_q$, as stated in the theorem below.

\begin{lemma}{\cite[Lemma 3]{SK08}}
Let $\mathcal{H}$ be the parity check matrix of an $[n,n-k]$ linear MRD code over $\mathbb{F}_{q^m}$. For any full rank $(n-k)\times n$ matrix $B$ over $\mathbb{F}_q$, the $n\times n$ matrix
$\begin{bmatrix}
   \mathcal{H}\\
    B \\
  \end{bmatrix}$
  is invertible.
\end{lemma}

Therefore, MRD codes will always achieve perfect secrecy irrespective of the network code used. The choice of the MRD code will only depend on the underlying field $\mathbb{F}_q$ of the network code.

\subsection{Byzantine Adversaries}

The malicious activity of the wiretapper in the model considered in this paper was restricted to eavesdropping. A more powerful wiretapper, with jamming capabilities, may not only listen to the data in the network but also alter it.
% Alex - removed - Due to the packet combining property of network codes,
This may lead to flooding the whole network with erroneous packets. Schemes to combat such wiretappers, known in literature as Byzantine adversaries, were studied in \cite{JLKHKMM08, KK08,SKK08} and the references within.

%
%Jaggi et al. considered in \cite{JLKHKMM08} three different scenarios where the wiretapper aims at preventing the transfer of information to the destinations by  jamming $\alpha$ links of his choice that are unknown to the destinations. Depending on the eavesdropping capabilities of the wiretapper, the authors distinguished between three possible scenarios, and  presented distributed polynomial algorithms with an exponentially decreasing probability of error that attain the  optimal multicast rate  for each of the following scenarios:
%\begin{enumerate}
%  \item The wiretapper is \emph{omniscient}, i.e. he has can observe all the links in the network. In this case, the optimal rate is $k=n-2\alpha$.
%  \item The wiretapper is omniscient with low rate secret channel  between the source and the destinations. In this case, the optimal rate is $k=n-\alpha$.
%  \item The wiretapper is \emph{limited}; he can eavesdrop to only $\mu$ links in the network with $2\alpha+\mu<n$. In this case, the optimal rate is also $k=n-\alpha$.
%\end{enumerate}
%
%In the previous three scenarios,  $k$ corresponds to the maximal rate at which the information can be sent reliably from the source to all the destinations in spite of the wiretapper interference. If, in addition, this information is to remain hidden from the wiretapper, the optimal rate may decrease dramatically. For instance, in the case of the omniscient wiretapper no information can be concealed from the wiretapper, and $k=0$.

 Consider a scenario where the wiretapper can not only observe $\mu$  edges but also jam $\alpha$ edges of his choice that are unknown to the destinations. In
%alex - can observed edges can also be jammed edge?
 this case, we will describe a coding scheme that achieves  a multicast rate of $k=n-2\alpha-\mu$ and guaranties that the information will remain hidden from the  wiretapper.   This can be achieved by  using a coset code as described in Section~\ref{sec:wtn} followed by a powerful network error-correcting code \cite{CY1,CY2}. First, we recall an important result in \cite[Theorem 4]{CY2}

\begin{theorem}
For an acyclic network $G(V,E)$ with min-cut $n$, there exists a linear $\alpha$-error-correcting code of dimension $(n-2\alpha)$ over a sufficiently large field.
\end{theorem}

 Let $\mathcal{G}$ be the generator matrix of a linear $\alpha$-error-correcting code of dimension $(n-2\alpha)$ whose existence is guaranteed by the previous theorem, and  Let $\mathcal{G}^\perp$ be its parity check matrix. A block diagram of the coding scheme that achieves secrecy against a Byzantine wiretapper at a rate $k=n-2\alpha-\mu$ is depicted  in Figure~\ref{fig:Biz}. First, the information $S=(s_1,\dots,s_k)^T$ is encoded using a coset code of parity check matrix $\mathcal{H}$ into the vector $T=(t_1,\dots,t_m)^T$, with $m=k+\mu$. The vector $T$ is then encoded into $Y=(y_1,\dots,y_n)^T=\mathcal{G}T$ using the network error-correcting code. To achieve perfect secrecy, $\mathcal{H}$ should satisfy the condition of  Theorem~\ref{th:oursec}, which can be expressed here as:
 \begin{equation}
\text{rank}
\begin{bmatrix}
    \mathcal{H} \\
    C_W\mathcal{G} \\
  \end{bmatrix}
  = k+\mu ~~ \text{for all
  $C_W$ s.t.} ~ \text{rank}(C_W)=\mu.
\label{eq:secc}
\end{equation}

We assume that the code is over a field large enough to guaranty the existence of the network error-correcting code and  the matrix $\mathcal{H}$ satisfying the above condition as well.
%alex - not clear - "as well as error correcting code
At each destination, a decoder  corrects the errors introduced by the wiretapper and recovers $T$. The information $S$ is then obtained as the unique solution of the system $\mathcal{H}S=T$. It was recently shown in \cite{NY09} that the  rate $k=n-2\alpha-\mu$ is optimal and another construction for codes with the same properties was presented there.

\section{Conclusion}\label{sec:conc}

We considered the problem of securing a multicast network
implementing network coding against a wiretapper capable of
observing a limited number of  edges of his choice, as defined initially by Cai and Yeung.
We showed that the problem can be formulated as a generalization of the wiretap channel of type II
which was introduced and studied by Ozarow and Wyner, and
decomposed into two sub-problems: the first one consists of designing a secure wiretap channel code, or a coset code, and the second consists of designing a network code satisfying some additional constraints. We proved there is no penalty to pay by adopting this separation, which we find in many ways illuminative.
%alex - we were not the first to prove it?
 Moreover, this approach allowed us to derive new bounds on the required alphabet size for secure codes. These new  bounds differ from those in the literature in that they are  independent from the network size and are functions of only the number of information symbols and that of destinations. We also analyzed the performance of the proposed coset codes under various wiretapper scenarios.
%alex - we can add that the codes are independend of the field size

A number of interesting questions related to this problem remain open. For instance, the bounds presented here on the code alphabet size can be large in certain cases and it is worthy to investigate whether  tighter bounds exist. Another  issue which was not addressed in this paper is that of designing efficient decoding algorithms at the destinations which can be very important
%alex - the decoding algorithm is pretty simple?
in practical implementations. Also, the work of \cite{SK08} hinted at some advantages of non-linear codes. The benefits of nonlinearity in security applications, whether at the source code or at the network code level, are still to be better understood.

\section*{Acknowledgments}
The authors would like to thank C.\ N.\ Georghiades for his continued support.
\bibliographystyle{ieeetran}{
\bibliography{ncsec}}
\end{document}